\documentclass[10pt]{article}
\usepackage{url}
\usepackage{setspace}
\usepackage{psfrag, graphicx, harvard, epsf,euscript,amsfonts,amsmath,latexsym,amssymb,latexsym}
\usepackage{amsthm}
\usepackage[active]{srcltx}
\usepackage{multicol}
\newcommand{\cS}{{\cal S}}
\newcommand{\cF}{{\cal F}}

\newcommand{\rE}{{\rm E}}

\newcommand{\rR}{\mathbb R}
\newcommand{\pr} {\par \noindent{\bf Proof\,:~}}
\newcommand{\epr}{\hfill\hbox{\hskip 4pt
                \vrule width 5pt height 6pt depth 1.5pt}\vspace{0.5cm}\par}
\newtheorem{lemma}{Lemma}
\newtheorem{theorem}{Theorem}
\newtheorem{assumption}{Assumption}
\newtheorem{remark}{Remark}
\newtheorem{definition}{Definition}

\def\noi{\noindent}
\title{{\bf Estimation of distribution functions in measurement error models}}
\author{
I. Dattner
\and
B.  Reiser
}
\begin{document}
\maketitle
\begin{abstract}
Many practical problems are related to the pointwise estimation of distribution functions when data contains measurement errors. Motivation for these problems comes from diverse fields such as astronomy, reliability, quality control, public health and survey data.
\par 
Recently, \citeasnoun{DGJ} showed that an estimator based on a direct inversion formula for distribution functions has nice properties when the tail of the  characteristic function of the measurement error distribution decays polynomially.  In this paper we derive theoretical properties for this estimator for the case where the error distribution is smoother and study its finite sample behavior for different error distributions. Our method is data-driven in the sense that we use only known information, namely, the error distribution and the data. Application of the estimator to estimating hypertension prevalence based on real data is also examined. 
\end{abstract}
\vspace*{1em}
\noi {\bf Keywords: adaptive estimator, deconvolution, error in variables, prevalence.}\\
\small
\normalsize
\section{Introduction}  
This research is motivated by the problem of pointwise estimation of distribution functions in the presence of measurement errors (distribution deconvolution). Interest in this problem goes back to \citeasnoun{eddington} who was motivated by astronomical data. \citeasnoun{gaffey} studied the problem of correcting for normal measurement errors in determining human cholesterol levels while \citeasnoun{Scheinok}, motivated by reliability theory, studied this problem under the assumption that the errors are exponentially distributed. In a quality control context, \citeasnoun{mee1984tolerance} studied the problem of estimating the proportion of a product satisfying a lower specification limit when the available data are subject to measurement error. Different approaches for estimating the finite population cumulative distribution function were developed for survey data, see \citeasnoun{Stefanski-Bay} and references therein. \citeasnoun{Nusser-Fuller} developed a semiparametric transformation approach to estimating usual daily intake distributions while \citeasnoun{Cordy-Thomas}, motivated by similar problems, suggested modeling the unknown distribution as a mixture of a finite number of known distributions. Also in the context of survey data, \citeasnoun{eltinge} develop adjusted estimators of distribution functions or quantiles for cases in which measurement errors are nonnormal. 
\par 
The methods developed in the papers cited above include both parametric and nonparametric approaches. Considering a nonparametric framework, the natural thing to do may be first to estimate the density and then integrating to obtain the estimator for the distribution function. This type of estimator was considered in \citeasnoun{zhang} and proved to be minimax optimal in \citeasnoun{Fan} for the case of supersmooth error distributions. However, \citeasnoun{Fan} was not able to show that this estimation method is optimal when the errors are ordinary smooth (e.g., double-exponential errors). We note that in the case of direct observations \citeasnoun{zhou} observed that optimality in density estimation does not carry over to distribution estimation. Recently, the case of ordinary smooth measurement errors was shown in \citeasnoun{DGJ} to be a more delicate one. In their work, a different estimation method was considered, namely, estimation based on a direct inversion formula for distribution functions. This deconvolution estimator was proved to be minimax optimal with no tail conditions being assumed for the estimated distribution (as has been required in all previous work). Also, based on Lepski's adaptation procedure (\citeasnoun{lepski}) they developed an adaptive algorithm for implementing the deconvolution estimator. 
\par 
In this paper we study further the problem of distribution deconvolution and consider both theoretical and practical aspects of the problem. The theoretical results are for the case of a known error distribution as is generally discussed in the deconvolution literature. In particular, the contribution of this research is as follows.
\begin{enumerate}
\item We show that a deconvolution estimator based on the direct inversion formula is minimax optimal also for supersmooth errors with no tail conditions being imposed on the estimated distribution. In addition, we develop the adaptive estimator for the supersmooth case and derive its statistical properties. 
\item We study the practical aspect of implementing the adaptive estimator through an extensive simulation study considering different error distributions and comparing it to the empirical distribution function and the SIMEX method. 
\item We apply the adaptive method to a real data example where one is interested in estimating hypertension prevalence in a population based on blood pressure measurements. 
\end{enumerate}
\par 
The rest of this paper is organized as follows. In section 2 we describe the estimation method and present the relevant theory for the supersmooth case. In section 3 we present the simulation study while in section 4 we apply our method to the real data example. A discussion follows in section 5 and proofs are provided in the appendix.

\section{The estimation method}\label{sec:addf}
\subsection{Deconvolution estimator}
The problem of estimating a distribution function in the presence of measurement errors is formulated mathematically as follows. Let $X_1, \ldots, X_n$ be a sequence of independent identically distributed random variables with common distribution $F_X$. Suppose that we observe random variables $Y_1, \ldots, Y_n$ given by
\begin{eqnarray}\label{eq:onesample-model}
Y_j=X_j+\epsilon_j,\ j=1,\ldots ,n,
\end{eqnarray}
where $\epsilon_j$ are independent identically distributed random variables, independent of $X_j$'s with a known density $f_\epsilon$ w.r.t. the Lebesgue measure on the real line.
Our objective is to estimate the cumulative distribution function $F_X(x_0)$
at any single given point $x_0\in \rR$ from the observations $Y_1, \ldots, Y_n$.
\par 
The deconvolution estimator presented in this paper is based on Fourier methods for which we introduce the following notation. Denote the characteristic function of a random variable $X$ by $\phi_X(\omega):=E e^{i\omega X}$, $\omega\in \rR$, and let $\Im(z)$ be the imaginary part of the complex variable $z$. Now, consider the inversion formula for a continuous distribution
(see \citeasnoun{gurland}, \citeasnoun{gil-pelaez} and
\citeasnoun[\S 4.3]{kendall}) 
\begin{equation}\label{eq:inversion}
F_X(x_0)=\frac{1}{2}-\frac{1}{\pi}\int_0^
\infty\frac{1}{\omega}
\Im\{e^{-i\omega x_0}\phi_X(\omega)\} d \omega,\;\;\;x_0\in \rR.
\end{equation} 
The above integral is interpreted as an improper Riemann integral. Assuming that $\phi_\epsilon$ is known, we use the fact that $\phi_X(\omega)=\phi_Y(\omega)/\phi_\epsilon(\omega)$, and replace $\phi_Y(\omega)$ by its empirical counterpart $\hat{\phi}_Y(\omega):=\frac{1}{n}\sum_{j=1}^ne^{i\omega Y_j}$. This leads to the following estimator for $F_X(x_0)$:
\begin{eqnarray}\label{eq:cdfEstimator}
\hat{F}_\lambda(x_0):=\frac{1}{2}-\frac{1}{\pi}
\int_0^\lambda\frac{1}{\omega}
\Im\bigg\{e^{-i\omega x_0}
\frac{\hat{\phi}_Y(\omega)}{\phi_\epsilon(\omega)}
\bigg\}d\omega,
\end{eqnarray}
where $\lambda>0$, is a predefined parameter (to be discussed later).
\par
This estimator is well defined if we assume that $|\phi_\epsilon(\omega)|\ne 0$ for all $\omega\in \rR$. This is a standard assumption in deconvolution problems; thus, throughout the paper we assume that the error characteristic function does not vanish. 
\begin{remark} 
In practice, the error distribution may not be completely known and additional information may be needed (e.g. repeated observations on $Y_j$ for a given $X_j$). In that case, a parametric approach may be taken for which the error distribution takes an explicit form (see below) depending on an unknown parameter for which an appropriate estimate may be used (we take this path when studying the real data example). A nonparametric approach would be to estimate $\phi_\epsilon$ and use it in the estimation procedure. We discuss this point in Section \ref{sec:dis}.  
\end{remark}
\par
We now take a deeper look into the deconvolution estimator (\ref{eq:cdfEstimator}). Generally, the estimator takes the form 
\begin{eqnarray}\nonumber 
\hat{F}_\lambda(x_0)&=&\frac{1}{2}-\frac{1}{n}\sum_{j=1}^n
I_\lambda(Y_j,x_0),
\\\label{eq:i-lambda}
I_\lambda(y,x_0)&:=&\frac{1}{\pi}\int_0^\lambda\frac{1}{\omega}
\Im\bigg\{
\frac{e^{i\omega(y-x_0)}}{\phi_\epsilon(\omega)}
\bigg\}d\omega.
\end{eqnarray}
Note that $I_\lambda(y,x_0)$ depends on the measurement error distribution. For example, in the case of Laplace error with zero expectation and scale parameter $\theta$ we have 
\[
I_\lambda(y,x_0)=
\frac{1}{\pi}\int_0^\lambda\frac{\sin\Big[\pi\omega(y-x_0)\Big]}{\omega}d\omega
+\frac{\theta^2\sin[\lambda(y-x_0)]}{\pi(y-x_0)^2}-\frac{\theta^2\lambda\cos[\lambda(y-x_0)]}{\pi(y-x_0)},
\]
while 
if the measurement error follows the normal distribution with standard deviation $\sigma_\epsilon$, then
\[
I_\lambda(y,x_0)=
\frac{1}{\pi}\int_0^\lambda\frac{\sin\Big[\omega(y-x_0)\Big]}{\omega}\exp\Big(\frac{\sigma_\epsilon^2\omega^2}{2}\Big)
d\omega.
\]
We see that the form of the deconvolution estimator is determined by the distribution of the measurement error. Lower bounds on rates of convergence show that the type of the error distribution is intrinsic to deconvolution problems. Indeed, it is well known that rates of convergence of the distribution/density function estimators in measurement error models are affected by the smoothness of the error density and the density to be estimated (see e.g. \citeasnoun{DGJ} and references therein). Smoothness is usually described by the tail behavior of the characteristic function, as in the following assumption for $\phi_\epsilon$ which characterizes supersmooth distributions. 
\begin{assumption}\label{as:global}
There exist positive constants $\beta> 0$, $\gamma>0$, $c_0>0$ and $c_1>0$ such that  
\begin{equation*}
c_0\exp(-\gamma|\omega|^\beta)\leq|\phi(\omega)|\leq c_1\exp(-\gamma|\omega|^\beta),\;\;\;
\forall\ \omega\in\rR.
\end{equation*}
\end{assumption}
The normal $(\beta=2)$ and Cauchy $(\beta=1)$ densities are
examples for which Assumption~\ref{as:global} holds.  In particular, the tails of the characteristic function of the normal and Cauchy decay exponentially. 
This is in contrast to the {\em ordinary smooth} case where the tail of $\phi_\epsilon$ decays in polynomial order. The spaces of ordinary smooth functions correspond to classic Sobolev classes, while supersmooth functions are infinitely differentiable. 
\par 
We also impose the following assumption. 
\begin{assumption}\label{as:local}
There exist positive real numbers 
$\omega_0$, $b_\epsilon$ and $\tau$ such that
\[
 |\phi_\epsilon(\omega)| \geq 1- b_\epsilon|\omega|^\tau,\;\;\;
\forall\ |\omega|\leq 
\omega_0.
\]
\end{assumption}
Assumption~\ref{as:local} describes the local behavior of the characteristic function of the error $\phi_\epsilon$ near the origin, and holds if $\phi_\epsilon$ is smooth at $\omega=0$. Since for any non--degenerate distribution there exist positive constants $b$ and $\delta$ such that 
$|\phi(\omega)|\leq 1- b|\omega|^2$ for all $|\omega|\leq \delta$ 
[see, e.g., \citeasnoun[Lemma~1.5]{petrov}], therefore we have $\tau\in (0, 2]$. 
\par 
We consider the Sobolev class of functions in order to express the smoothness of the estimated distribution $F_X$.
\begin{definition}
Let $\alpha >-1/2$, $L>0$. We say that $F_X$ belongs to the class $\cS_\alpha(L)$
if it has a density $f_X$ with respect to the Lebesgue measure, and
\[
\frac{1}{2\pi}
\int_{-\infty}^\infty|\phi_X(\omega)|^2 (1+\omega^2)^{\alpha}\,
d \omega\leq L^2.
\]
\end{definition}
The set $\cS_\alpha(L)$ with $\alpha>-1/2$ contains absolutely
continuous distributions while if $\alpha>1/2$ then $\cS_\alpha(L)$ contains distributions with bounded continuous densities. 
\par 
In our study of the rates of convergence of the deconvolution estimator we bound the maximal (pointwise root mean squared error) risk of the estimator over the nonparametric family $\cS_\alpha(L)$ defined above. Rates of convergence of the estimator (\ref{eq:cdfEstimator}) for the case of ordinary smooth error and $F_X\in \cS_\alpha(L)$ were studied in \citeasnoun{DGJ}. The following theorem establishes rates of convergence for the supersmooth case. %
\begin{theorem}\label{th:supersmooth}
Let the observations be given by model (\ref{eq:onesample-model}). Let the estimator for $F_X(x_0)$ be $\hat{F}_{\lambda}(x_0)$ as defined in~(\ref{eq:cdfEstimator}) and associate with the parameter
\[
\lambda=\lambda_\star:=
\Big\{\frac{\ln n}{2\gamma}-\frac{\ln c_\epsilon+\frac{(2\alpha+2)}{\beta}\ln\Big(\frac{\ln n}{2\gamma}\Big)-2\ln(K_0L)}
{2\gamma}
\Big\}^{1/\beta}.
\]
If $\alpha>-1/2$ and Assumptions~\ref{as:global}-\ref{as:local} hold, then we have for all $x_0\in \rR$ and large enough $n$ 
\begin{equation}\label{eq:minimax}
\sup_{F_X\in \cS_\alpha(L)}\Big\{E|\hat{F}_{\lambda_\star}(x_0)-F_X(x_0)|^2\Big\}^{1/2}\leq
K_0L\Big(\frac{\ln n}{4\gamma}\Big)^{-(\alpha+1/2)/\beta},
\end{equation}
where $K_0:=\sqrt{2/\pi}[1+ (2\alpha+1)^{-1/2}]$ and $c_\epsilon$ depends only on the error distribution and is defined in (\ref{eq:ceps}).
\end{theorem}
\par
Unlike the case of ordinary smooth errors the rate of convergence in the supersmooth case is very slow, logarithmic in the sample size $n$. We note that this rate of convergence is minimax optimal for $\alpha>1/2$. In order to prove such a result one needs to show that the maximal risk~(\ref{eq:minimax}) matches up to a constant the minimal attainable risk for this problem. Indeed, under additional standard assumptions on $\phi_\epsilon$ it can be shown that if $\alpha>1/2$ and the class $\cS_\alpha(L)$ is rich enough, then without loss of generality we have for all $n$ large enough
\[
\inf_{\hat{F}_X} \sup_{F_X\in \cS_\alpha(L)}\Big\{E|\hat{F}_X(0)-F_X(0)|^2\Big\}^{1/2}\geq
C[\ln n]^{-(\alpha+1/2)/\beta},
\]
where $C$ is a positive constant independent of $n$ and $\inf$ is taken over all possible estimators $\hat{F}_X(0)$ of $F_X(0)$. This lower bound on the minimax risk is in the same order as the upper bound given in Theorem \ref{th:supersmooth}. Thus, the estimator (\ref{eq:cdfEstimator}) with the choice $\lambda=\lambda_\star$ is optimal in order. That is to say that no other estimator can do better (in the minimax sense). This result can be proved in the same way \citeasnoun{DGJ} derived the lower bound for the case of ordinary smooth errors. Under additional assumptions on the tail behavior of $F_X$, \citeasnoun{Fan} derived minimax optimal rates of convergence for estimation over H\"older classes.
\par 
The optimal choice of the parameter $\lambda=\lambda_\star$ as given in the theorem is a result of the standard bias-variance trade-off. The bias of the estimator depends only on the distribution of $X$ and decreases as $\lambda$ increases. On the other hand, the variance is affected by the tail behavior of the error characteristic function $\phi_\epsilon$ and is increasing with $\lambda$. It is clear that the role of the design parameter $\lambda$ is crucial. The problem is that in practice we do not know the value of the class parameters $\alpha$, $L$ and therefore $\lambda_\star$ as defined in the theorem can not be calculated. In the next section we show how to choose the "bandwidth" parameter $\lambda$ based only on the information we have, namely, the given data and the assumed error distribution. 
\subsection{Adaptive deconvolution estimator}
We first develop an adaptive version of the estimator for the case of supersmooth error distribution and provide its theoretical properties. Then we 
discuss the ordinary smooth case were we mimic the optimal choice $\lambda=\lambda_\star$ by an adaptive algorithm based on Lepski's adaptation procedure~\cite{lepski}. The theoretical properties of the resulting estimator in the ordinary smooth case were studied in \citeasnoun{DGJ} who showed that the adaptive estimator is consistent and achieves the optimal rate of convergence within a logarithmic factor (it can be shown that the logarithmic factor cannot be eliminated, see \citeasnoun{lepski}). 
\par 
We now develop an adaptive version of the estimator for the case of supersmooth error. In particular, the next theorem shows that there is no additional payment for adaption in this case.
\begin{theorem}\label{th:supersmooth-adaption}
Let the observations be given by model (\ref{eq:onesample-model}). Let the estimator for $F_X(x_0)$ be $\hat{F}_{\lambda}(x_0)$ as defined in~(\ref{eq:cdfEstimator}) and associate with the parameter
\[\lambda=\hat{\lambda}:=
\Big\{\frac{\ln n}{2\gamma}-\frac{\ln c_\epsilon+\Big[\ln\Big(\frac{\ln n}{2\gamma}\Big)\Big]^2}
{2\gamma}
\Big\}^{1/\beta}.
\]
If $\alpha>-1/2$ and Assumptions~\ref{as:global}-\ref{as:local} hold, then we have for all $x_0\in \rR$ and large enough $n$ 
\[
\sup_{F_X\in \cS_\alpha(L)}\Big\{E|\hat{F}_{\hat{\lambda}}(x_0)-F_X(x_0)|^2\Big\}^{1/2}\leq
K_0L\Big(\frac{\ln n}{4\gamma}\Big)^{-(\alpha+1/2)/\beta},
\]
where $K_0:=\sqrt{2/\pi}[1+ (2\alpha+1)^{-1/2}]$ and $c_\epsilon$ depends only on the error distribution and is defined in (\ref{eq:ceps}).
\end{theorem}
 

Note that the rate of convergence in the theorem is the optimal one when $\alpha>1/2$. Moreover, $\hat{\lambda}$ does not depend on the class parameters $\alpha$ and $L$. In particular, $\hat{\lambda}$ is smaller than $\lambda_\star$ (as defined in Theorem~\ref{th:supersmooth}) which depends on $\alpha$ in a term of second order. Therefore, the small modification of $\lambda_\star$ which makes the bias dominant in the bias-variance trade off, does not affect the rate of convergence.  
\par 
We now turn to the case of ordinary smooth error distribution. Consider the set of positive parameters $\Lambda:=\{\lambda_{\min},\ldots,\lambda_{\max}\}$, and the family of estimators
$\cF_\Lambda:=\big\{\hat{F}_\lambda(x_0), \lambda\in \Lambda\big\}$, where  
$\hat{F}_\lambda(x_0)$ is given by (\ref{eq:cdfEstimator}). Define
\begin{equation}\label{eq:sigma_lambda}
\hat{\sigma}_\lambda:=\Big[
\frac{1}{n}\sum_{j=1}^n\{I_\lambda(Y_j,x_0)\}^2\Big]^{1/2},
\end{equation}
where $I_\lambda$ is given by (\ref{eq:i-lambda}). The adaptive estimator $\hat{F}_A(x_0)$ is obtained 
by selecting from the family $\cF_\Lambda$
according to the following rule. Let $K_\epsilon=0.0275+0.3074\sigma_\epsilon$, and with any estimator
$\hat{F}_\lambda(x_0)$ we associate the interval
\begin{equation*}
Q_\lambda:= \Big[\hat{F}_\lambda(x_0) - 
K_\epsilon\Big\{\frac{\ln(n)}{n}\Big\}^{1/2}\hat{\sigma}_\lambda,\;
\hat{F}_\lambda(x_0) + K_\epsilon\Big\{\frac{\ln(n)}{n}\Big\}^{1/2}\hat{\sigma}_\lambda\Big],
\end{equation*}
and define
\begin{equation}\label{eq:adaptest}
\hat{F}_A(x_0):=\hat{F}_{\hat{\lambda}}(x_0),
\end{equation}
where
\begin{equation*}
 \hat{\lambda}:=\min\bigg\{\lambda\in \Lambda: \bigcap_{
\mu\geq\lambda,\, \mu\in \Lambda} Q_\mu \;\ne \;\emptyset\bigg\}.
\end{equation*}
We use below the set $\Lambda=0.01(0.05)10$ and the projection of $\hat{F}_A(x_0)$ on the interval $[0,1]$ as the final estimator.
\par 
The value of $K_\epsilon$ as specified above is a result of the tuning of the adaptive algorithm. Although according to the theory, for a given error distribution one can determine the constant $K_\epsilon$, it turns out to be too conservative in practice. This problem was already noted by \citeasnoun{spokoiny} who proposed a tuning approach for a different model. 
\par
A detailed explanation of our tuning approach is given in the appendix. We note that we "tuned" our algorithm according to the Laplace error. In the sequel we use this rule for all error distributions including the normal one (and not the adaptive estimator defined in Theorem~\ref{th:supersmooth-adaption} for the supersmooth case). Ideally, we could calibrate our estimator specifically for a given error distribution. However, considering the long computational time of calibration and the fact that the performance of the adaptive estimator in simulations does not seem to be very sensitive to this assumption, we use this rule for all measurement error models in our simulation study. 

\section{Simulation study}\label{sec:sim}
\subsection{Study description}
The following set up is used in our simulation study. The unobserved distribution $F_X$ is assumed to be one of the following. 
\begin{enumerate}
\item Gamma with shape parameters $3$ and scale $1/\sqrt{3}$.
\item Standard normal.
\end{enumerate}
Define the standard deviations of $X$ and $\epsilon$ by $\sigma_X$ and $\sigma_\epsilon$ respectively. The error distributions are chosen such that we have a specific noise to signal ratio $\sigma_\epsilon/\sigma_X$. In particular, we are interested in the values $\sigma_\epsilon/\sigma_X=0.2,\ 0.5$, corresponding to $20\%,\ 50\%$ error contamination respectively. We consider eight error distributions as follows.
\begin{enumerate}
\item Gamma distribution with shape parameter two, and scale parameters $\theta=1/(5\sqrt{2}),\ 1/(2\sqrt{2})$.
\item As in $(1)$ but relocated to have zero expectation.
\item Laplace distribution with zero expectation and the same scale parameters as in $(1)$.
\item Normal distribution with zero expectation and standard deviations $\sigma_\epsilon=1/5,\ 1/2$.
\end{enumerate}
Two of the above ($3.$ and $4.$) provide error distributions which are symmetric around zero but differ in their tail properties. The other two are skewed distributions with $(1.)$ resulting in only positive values while $(2.)$ allows for negative values as well.
\par 
Usually, measurement errors are considered to have zero expectation but in some cases this appears not to hold. In the context of blood pressure \citeasnoun{marshall} discusses that the presence of a medical student results in an increase in measured  blood pressure. \citeasnoun{walker} in a robustness study of ANOVA consider a beta distribution with nonzero expectation as a possible model for measurement errors. \citeasnoun{albers} in the context of screening production processes discuss situations with nonzero expectation for measurement error. 
\par
All together, we have sixteen combinations of measurement error models. Each combination is simulated for sample sizes $n=100$, and $500$, resulting in thirty two different experimental set ups. For each experimental set up, $1000$ independent samples of size $n$ were generated, from which we estimated for various values of $x_0$, $F_X(x_0)$ where $x_0$ values were chosen to correspond to the percentiles $0.1,0.25,0.5,0.75,0.9$ of the unobserved distribution $F_X$.
\par
In all the scenarios just defined, the behavior of the adaptive estimator (\ref{eq:adaptest}) was compared to two other estimators. The first is the empirical distribution function of the observations which we call the naive estimator,
\[\hat{F}_Y(x_0):=\frac{1}{n}\sum_{j=1}^n\textbf{1}(Y_j\leq x_0),\] where $\textbf{1}(\cdot)$ stands for the indicator function. The second is the SIMEX (simulation extrapolation) estimator $\hat{F}_S(x_0)$ introduced in \citeasnoun{Stefanski-Bay}, which we describe now. 
\par
In simulation extrapolation, estimators are recomputed on a large number $B$ of measurement error-inflated, pseudo data sets, $\{Y_{j,b}(\tau)\}_{j=1}^n$, $(b=1,...,B)$, with 
\[Y_{j,b}(\tau)=Y_j+\sqrt{\tau}\epsilon^*_{j,b},\ (j=1,...,n,\ b=1,...,B),\]
where $\epsilon^*_{j,b}\sim f_\epsilon$ are independent, pseudo-random variables and $\tau\geq 0$ is a constant controlling the amount of added error. According to this setup the total measurement error variance in $Y_{j,b}(\tau)$ is $\sigma_\epsilon^2(\tau+1)$. Thus, the general idea is based on the fact that if we let $\tau=-1$ then we end up with zero measurement error in the random variables $Y_{j,b}(\tau)$ . 
\par 
The cumulative distribution function estimator calculated from the $b$th variance-inflated data set $Y_{j,b}(\tau)$ is called the $b$th pseudo estimator, and is
\[\frac{1}{n}\sum_{j=1}^n\textbf{1}(Y_{j,b}(\tau)\leq x_0),\ (b=1,...,B).\]
We now average the pseudo estimators and define
\[\hat{F}_{Y,\tau,n}(x_0)=\frac{1}{B}\sum_{b=1}^B\Big[\frac{1}{n}\sum_{j=1}^n\textbf{1}(Y_{j,b}(\tau)\leq x_0)\Big].\]
The SIMEX method is based on the assumption that the expectation $E[\hat{F}_{Y,\tau,n}(x_0)]$ can be well approximated by a quadratic function of $\tau$: $\beta_0+\beta_1\tau+\beta_2\tau^2$, for constants $\beta_0,\beta_1,\beta_2$ depending on $x_0$, $\sigma_\epsilon^2$ and $F_X$. For a given sequence $\tau_1,...,\tau_m$, the SIMEX procedure require to estimate $\{\hat{F}_{Y,\tau_1,n}(x_0),...,\hat{F}_{Y,\tau_m,n}(x_0)\}$, so that $\beta_0,\beta_1,\beta_2$ can be estimated by a least squares regression of $\{\hat{F}_{Y,\tau_1,n}(x_0),...,\hat{F}_{Y,\tau_m,n}(x_0)\}$ on $\tau_1,...,\tau_m$, yielding the estimates $\hat{\beta}_0,\ \hat{\beta}_1,\ \hat{\beta}_2$. Extrapolation to the case of no measurement errors is accomplished by letting $\tau\rightarrow-1$, resulting in the SIMEX estimator 
\[\hat{F}_S(x_0):=\hat{\beta}_0-\hat{\beta}_1+\hat{\beta}_2.\]
In our simulations $B=2000$ and following \citeasnoun{Stefanski-Bay} we set $\tau=0.05(0.4875)2$.
\subsection{Numerical results}
Tables~\ref{tab:est_normal1}-\ref{tab:est_gamma2} summarize the empirical root mean square error and bias of the three estimators described above for the different experimental set ups. We present only the results for sample size $n=500$, since they are similar to those for $n=100$, but are more stable. For each error distribution in the tables, the first block is for $20\%$ contamination while the second block is for $50\%$ contamination. The observed absolute value of the bias$\times$10 of the estimator is given in parentheses. 
\par
In Tables~\ref{tab:est_normal1} and \ref{tab:est_gamma1} we see that when the error takes only positive values, i.e., is Gamma distributed, then the adaptive estimator achieves better results uniformly over the distribution of $X$ for both $20\%$ and $50\%$ contamination. The bias of the SIMEX and naive estimators is very large in these cases. When the distribution of the error is Gamma around zero, then the performance of the SIMEX and naive estimators substantially improves. However, the adaptive estimator is usually better in root mean square error, and when not, its root mean square error value is close to the best.
\par
For Laplace distributed measurement error the results are similar for both $X$ distributions. When the contamination is $20\%$ the adaptive estimator is again uniformly better than the other two. However, the results are more mixed when we have $50\%$ contamination. 
\par 
When the error is normally distributed, the results are mixed. Here, the root mean square error of the adaptive estimator is high when estimating lower and upper quantiles under $20\%$ contamination, but has the same order as SIMEX for estimating other quantiles. Note that in terms of root mean square error, the naive estimator performs very well under normal error with small contamination. \begin{remark}
Recalling that for normal error the optimal minimax rates are very slow (logarithmic in the sample size), one may wonder how in practice the estimation results seems to be reasonable as implied by our simulation study. This may be a result of the essentially small error variance, see for example \citeasnoun{Fan2} who  studied how large a noise level is acceptable under supersmooth error distributions.
\end{remark}

\begin{table}
\caption{Empirical root mean square error and bias$\times$10 (in parenthesis) for estimating standard normal under non symmetric error distribution.}																					
\label{tab:est_normal1}	
\begin{center}
\begin{tabular}{lccccc}																					
\hline
\hline																					
\multicolumn{1}{l}{}&																					
\multicolumn{5}{c}{$F_X(x_0)$}\\																					
\hline																					
Estimator&0.1&0.25&0.5&0.75&0.9\\																					
											
\hline		
\\																			
\multicolumn{6}{c}{Gamma error - $20\%$ contamination}\\\\																		
Adaptive	&	0.013 (0.031)	&	0.020 (0.032)	&	0.022	(0.004)&	0.019	(0.028)&	0.013	(0.014)		\\
SIMEX	&	0.020	(0.119)&	0.029	(0.158)&	0.031 (0.056)	&	0.032	(0.149)&	0.032	(0.229)	\\
Naive	&	0.039	(0.373)&	0.078	(0.756)&	0.111	(1.093)&	0.102	(1.002)&	0.065	(0.630)	\\\\
\multicolumn{6}{c}{Gamma error - $50\%$ contamination}\\\\													
Adaptive	&	0.019	(0.042)&	0.026	(0.051)&	0.027	(0.019)&	0.028	(0.037)&	0.024	(0.044)	\\
SIMEX	&	0.048	(0.458)&	0.087	(0.833)&	0.097	(0.908)&	0.040	(0.146)&	0.073	(0.676)		\\
Naive	&	0.066	(0.656)&	0.145	(1.442)&	0.237	(2.361)&	0.254	(2.528)&	0.198	(1.966)		\\\\
\multicolumn{6}{c}{Gamma error with zero expectation - $20\%$ contamination}\\\\																					
Adaptive	&	0.013	(0.020)&	0.019	(0.031)&	0.021	(0.003)&	0.019	(0.028)&	0.014	(0.024)	\\
SIMEX	&	0.016	(0.005)&	0.023	(0.008)&	0.027	(0.001)&	0.022	(0.004)&	0.016	(0.001)	\\
Naive	&	0.014	(0.039)&	0.020	(0.051)&	0.023	(0.003)&	0.020	(0.042)&	0.014	(0.046)	\\\\
\multicolumn{6}{c}{Gamma error with zero expectation - $50\%$ contamination}\\\\															
Adaptive	&	0.018	(0.035)&	0.026	(0.050)&	0.028	(0.001)&	0.030	(0.056)&	0.024	(0.044)		\\
SIMEX	&	0.020	(0.007)&	0.027	(0.035)&	0.031	(0.045)&	0.027	(0.023)&	0.021	(0.001)	\\
Naive	&	0.027	(0.232)&	0.033	(0.263)&	0.024	(0.077)&	0.026	(0.175)&	0.030	(0.256)	\\\\
\hline	
\end{tabular}																					
\end{center}																					
\end{table}

\begin{table}		
\caption{Empirical root mean square error and bias$\times$10 (in parenthesis) for estimating Gamma with shape three and scale $1/\sqrt{3}$ under non symmetric error distribution.}
\label{tab:est_gamma1}																			
\begin{center}																					
\begin{tabular}{lccccc}																					
\hline	
\hline																				
\multicolumn{1}{l}{}&																					
\multicolumn{5}{c}{$F_X(x_0)$}\\																					
\hline																					
Estimator&0.1&0.25&0.5&0.75&0.9\\																					
\hline																					
	\\																				
\multicolumn{6}{c}{Gamma error - $20\%$ contamination}\\	\\																				
Adaptive	&	0.014 (0.041)	&	0.018	(0.003)&	0.023	(0.021)&	0.019	(0.001)&	0.014	(0.013)		\\
SIMEX	&	0.045	(0.420)&	0.041	(0.321)&	0.034	(0.084)&	0.037	(0.234)&	0.026	(0.154)		\\
Naive	&	0.065	(0.642)&	0.112	(1.113)&	0.128	(1.264)&	0.092	(0.893)&	0.046	(0.434)	\\\\
\multicolumn{6}{c}{Gamma error - $50\%$ contamination}\\	\\																					
Adaptive	&	0.021	(0.056)&	0.027	(0.032)&	0.032	(0.057)&	0.029	(0.036)&	0.021	(0.014)	\\
SIMEX	&	0.087	(0.871)&	0.161	(1.600)&	0.137	(1.332)&	0.048	(0.296)&	0.093	(0.917)	\\
Naive	&	0.088	(0.883)&	0.190	(1.896)&	0.281	(2.801)&	0.252	(2.512)&	0.150	(1.482)	\\\\

\multicolumn{6}{c}{Gamma error with zero expectation - $20\%$ contamination}\\	\\																				
Adaptive	&	0.014	(0.041)&	0.018	(0.003)&	0.023	(0.021)&	0.019	(0.001)&	0.014	(0.013)		\\
SIMEX	&	0.019	(0.007)&	0.025	(0.006)&	0.027	(0.008)&	0.022	(0.003)&	0.016	(0.006)		\\
Naive	&	0.018	(0.108)&	0.020	(0.047)&	0.023	(0.027)&	0.020	(0.049)&	0.014	(0.032)		\\\\
\multicolumn{6}{c}{Gamma error with zero expectation - $50\%$ contamination}\\	\\																								
Adaptive	&	0.021	(0.051)&	0.026	(0.030)&	0.033	(0.059)&	0.030	(0.030)&	0.021	(0.013)	\\
SIMEX	&	0.025	(0.094)&	0.030	(0.073)&	0.031	(0.021)&	0.026	(0.002)&	0.019	(0.001)	\\
Naive	&	0.053	(0.509)&	0.037	(0.309)&	0.024	(0.054)&	0.031	(0.247)&	0.024	(0.192)	\\\\
			\hline																		
\end{tabular}																					
\end{center}																					
																					
\end{table}

\begin{table}
\caption{Empirical root mean square error and bias$\times$10 (in parenthesis) for estimating standard normal under symmetric error distribution.}																					
\label{tab:est_normal2}	
\begin{center}
\begin{tabular}{lccccc}																					
\hline
\hline																					
\multicolumn{1}{l}{}&																					
\multicolumn{5}{c}{$F_X(x_0)$}\\																					
\hline																					
Estimator&0.1&0.25&0.5&0.75&0.9\\																					
											
\hline		
\\																										
\multicolumn{6}{c}{Laplace error - $20\%$ contamination}\\		\\																			
Adaptive	&	0.013	(0.027)&	0.019	(0.012)&	0.021	(0.002)&	0.019	(0.027)&	0.013	(0.010)	\\
SIMEX	&	0.016	(0.007)&	0.023	(0.013)&	0.026	(0.001)&	0.022	(0.003)&	0.016	(0.014)\\
Naive	&	0.014	(0.051)&	0.020	(0.029)&	0.023	(0.001)&	0.019	(0.043)&	0.014	(0.032)	\\\\
\multicolumn{6}{c}{Laplace error - $50\%$ contamination}\\		\\																							
Adaptive	&	0.022	(0.055)&	0.027	(0.047)&	0.029	(0.003)&	0.029	(0.044)&	0.022	(0.044)		\\
SIMEX	&	0.019	(0.005)&	0.025	(0.003)&	0.029	(0.002)&	0.026	(0.001)&	0.020	(0.009)	\\
Naive	&	0.029	(0.253)&	0.028	(0.210)&	0.023	(0.004)&	0.029	(0.211)&	0.029	(0.243)		\\\\
\multicolumn{6}{c}{Normal error - $20\%$ contamination}\\	\\																				
Adaptive	&	0.032	(0.286)&	0.022	(0.128)&	0.019	(0.005)&	0.023	(0.138)&	0.032	(0.290)	\\
SIMEX	&	0.016	(0.005)&	0.023	(0.002)&	0.025	(0.005)&	0.024	(0.005)&	0.016	(0.000)	\\
Naive	&	0.015	(0.051)&	0.020	(0.045)&	0.021	(0.004)&	0.021	(0.040)&	0.014	(0.042)	\\\\
\multicolumn{6}{c}{Normal error - $50\%$ contamination}\\	\\																									
Adaptive	&	0.025	(0.186)&	0.029	(0.198)&	0.019	(0.003)&	0.030	(0.210)&	0.024	(0.180)	\\
SIMEX	&	0.020	(0.012)&	0.027	(0.008)&	0.031	(0.001)&	0.027	(0.028)&	0.020	(0.014)		\\
Naive	&	0.030	(0.260)&	0.030	(0.225)&	0.023	(0.001)&	0.031	(0.237)&	0.030	(0.262)		\\\\
\hline															
\end{tabular}																					
\end{center}																					
																				
\end{table}

\begin{table}		
\caption{Empirical root mean square error and bias$\times$10 (in parenthesis) for estimating Gamma with shape three and scale $1/\sqrt{3}$ under symmetric error distribution.}
\label{tab:est_gamma2}																			
\begin{center}																					
\begin{tabular}{lccccc}																					
\hline	
\hline																				
\multicolumn{1}{l}{}&																					
\multicolumn{5}{c}{$F_X(x_0)$}\\																					
\hline																					
Estimator&0.1&0.25&0.5&0.75&0.9\\																					
\hline																					
	\\																										
\multicolumn{6}{c}{Laplace error - $20\%$ contamination}\\		\\																			
Adaptive	&	0.014	(0.028)&	0.019	(0.004)&	0.021	(0.020)&	0.019	(0.018)&	0.014	(0.017)	\\
SIMEX	&	0.018	(0.014)&	0.024	(0.003)&	0.025	(0.006)&	0.021	(0.016)&	0.015	(0.001)	\\
Naive	&	0.017	(0.085)&	0.020	(0.028)&	0.022	(0.032)&	0.019	(0.032)&	0.014	(0.024)	\\\\
\multicolumn{6}{c}{Laplace error - $50\%$ contamination}\\		\\																							
Adaptive	&	0.026	(0.056)&	0.029	(0.022)&	0.033	(0.055)&	0.027	(0.011)&	0.019	(0.010)\\
SIMEX	&	0.022	(0.022)&	0.027	(0.024)&	0.030	(0.024)&	0.026	(0.029)&	0.018	(0.009)	\\
Naive	&	0.045	(0.423)&	0.027	(0.177)&	0.027	(0.141)&	0.031	(0.232)&	0.022	(0.168)	\\\\
\multicolumn{6}{c}{Normal error - $20\%$ contamination}\\	\\																				
Adaptive	&	0.029	(0.257)&	0.023	(0.137)&	0.021	(0.003)&	0.022	(0.110)&	0.030	(0.273)		\\
SIMEX	&	0.019	(0.003)&	0.023	(0.000)&	0.027	(0.003)&	0.023	(0.010)&	0.015	(0.005)	\\
Naive	&	0.017	(0.099)&	0.019	(0.029)&	0.023	(0.039)&	0.020	(0.038)&	0.014	(0.023)		\\\\
\multicolumn{6}{c}{Normal error - $50\%$ contamination}\\	\\																								
Adaptive	&	0.040	(0.357)&	0.027	(0.168)&	0.030	(0.197)&	0.029	(0.198)&	0.016	(0.060)	\\
SIMEX	&	0.026	(0.109)&	0.028	(0.002)&	0.031	(0.073)&	0.027	(0.014)&	0.019	(0.001)	\\
Naive	&	0.052	(0.493)&	0.029	(0.206)&	0.028	(0.180)&	0.034	(0.273)&	0.022	(0.170)		\\\\
			\hline																		
\end{tabular}																					
\end{center}																					
																					
\end{table}																					
		
\par
\begin{figure}
\centerline{\includegraphics[width=5in]{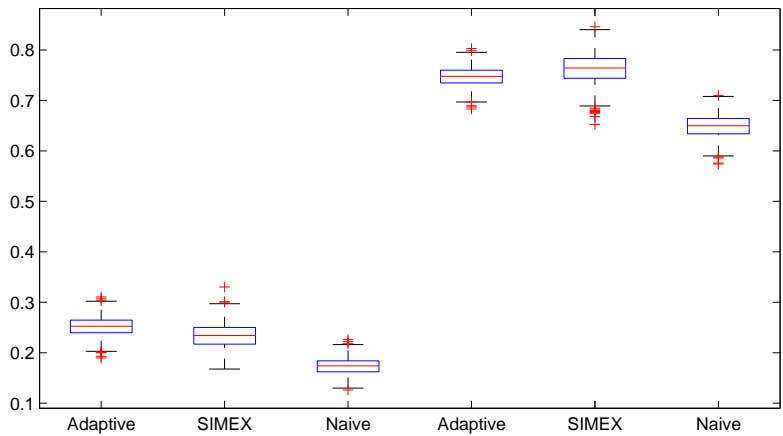}}
\caption{The effect of the shape of the error distribution on the performance of the estimators. Here $X\sim N(0,1)$, $\epsilon$ is Gamma distributed with shape parameter two and scale parameter $1/(5\sqrt{2})$, MC=$1000$ and $n=500$.}
\label{fig:shape}\end{figure}

Summarizing the numerical results, we see that the adaptive estimator performs reasonably well regardless of the shape and location of the error distribution while the SIMEX and naive estimators do not. Indeed, when the error is Gamma distributed, there are cases where the empirical root mean square error of the adaptive estimator is about one tenth of the empirical root mean square error of the naive estimator. This phenomenon is illustrated in Figure~\ref{fig:shape}. We present there box plots for the case where $X\sim N(0,1)$ and $\epsilon$ is Gamma distributed with shape parameter two and scale parameter $1/(5\sqrt{2})$ over the $1000$ Monte Carlo simulations based on a sample size of $n=500$. In the figure we focus on the estimation of the cumulative probabilities $0.25$ and $0.75$. The box plots for the adaptive, SIMEX and naive estimator are displayed side by side. It is clear from the plots that the naive estimator is totally wrong for the asymmetric error distribution. The SIMEX is less affected and the adaptive estimator achieves the best result. When the measurement error distribution is symmetric, the results are mixed with no method being superior all the time. However, we note that for larger sample sizes, we expect the naive estimator to be worse than the adaptive estimator since the naive estimator is not consistent. 

\par 
MATLAB code for executing all simulations described above and implementing the adaptive estimator for user data is available at \url{http://stat.haifa.ac.il/~idattner/add}.
\section{Estimating hypertension prevalence}
\subsection{Data description}
High blood pressure (hypertension) is a direct cause of serious cardiovascular disease (\citeasnoun{kannel}) and estimating hypertension prevalence is of substantial interest. Specifically, a blood pressure level of $140/90$ mmHg or greater is considered high. However, blood pressure is known to be measured with additional error which needs to be addressed in its analysis (see e.g., \citeasnoun{marshall} and references therein). Thus, treating the observed blood pressure measurements naively and estimating hypertension prevalence with, say, the empirical distribution function, would result in a biased estimate.  
\par
We illustrate our method using data from the Framingham Heart Study (\citeasnoun{carroll-et-al}). This study consists of a series of exams taken two years apart. We use systolic blood pressure (SBP) measurements of $1,615$ men aged $31-65$, from Exam two and Exam three. We treat the SBP values of each individual $j$ for the two exams ($Y_{j,1}$, $Y_{j,2}$) as repeated measures of the long-term average SBP, which is denoted by $X_j$:
\begin{eqnarray}\label{eq:repeatedmodel}
Y_{j,1}&=&X_j+\epsilon_{j,1},
\\\nonumber
Y_{j,2}&=&X_j+\epsilon_{j,2},
\end{eqnarray}
for individuals $j=1,...,n$. 
\par
\begin{figure}
\centerline{\includegraphics[width=4in]{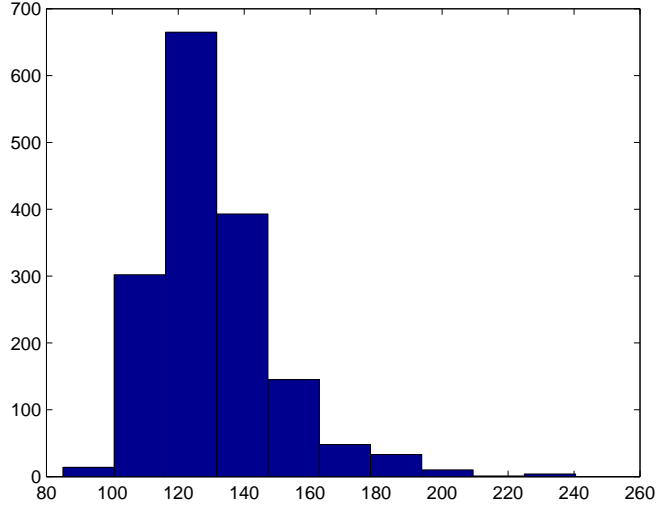}}
\caption{Systolic blood pressure measurements of $1,615$ men aged $31-65$ from the Framingham Heart Study.}
\label{fig:fram}
\end{figure}

 
Following \citeasnoun{carroll-et-al}, we use the average of the two exams $Y^{\prime}_j=(Y_{j,1}+Y_{j,2})/2$, so that the model in our case is
\begin{eqnarray}\label{eq:realmodel}
Y^{\prime}_j=X_j+\epsilon^{\prime}_j,
\end{eqnarray}
where $\epsilon^{\prime}_j=(\epsilon_{j,1}+\epsilon_{j,2})/2$, and we are interesting in the estimation of $1-F_X(140)$ from the data $Y^{\prime}_j,\ j=1,...,1615$. An histogram of the data $Y^{\prime}$ is displayed in Figure~\ref{fig:fram}.

\par 
Note that the repeated measures model (\ref{eq:repeatedmodel}) represents a balanced random effects model, thus the measurement error variance estimate (\citeasnoun{searle}) is
\begin{eqnarray}\label{eq:varest}
\hat{\sigma}_\epsilon^2&=&\sum_{j=1}^n\sum_{k=1}^p\frac{(Y_{j,k}-\bar{Y}_{j.})^2}{n(p-1)},
\end{eqnarray}
where $\bar{Y}_{j.}:=\frac{1}{p}\sum_{k=1}^pY_{j,k}$ is the sample mean for each individual $j$. In our case $n=1,615$, $p=2$ and the measurement error variance estimate is $\hat{\sigma}_\epsilon^2=84.755$.
\par 
An important aspect in the model described here that we did not consider in our simulation study of Section~\ref{sec:sim} is that $\sigma_\epsilon$ is not known but estimated from the data. In order to understand how this practical feature affects our method, we performed another simulation study, based on the model as defined in (\ref{eq:repeatedmodel})-(\ref{eq:realmodel}), in which we assume that $\epsilon\sim N(0,9.206^2)$ and $X\sim N(130.757,17.528^2)$. In particular, the simulation step of the SIMEX estimator is based on $\hat{\sigma}_\epsilon^2$ as given by (\ref{eq:varest}) and our method is based on a standardized version of (\ref{eq:realmodel}), i.e.,  $(Y^{\prime}_j-\frac{1}{n}\sum_{j=1}^nY^{\prime}_j)/\hat{\sigma}_{Y^{\prime}}$ and the estimated variance $\hat{\sigma}_\epsilon/\hat{\sigma}_{Y^{\prime}}$ (the standardization is needed because of the way we tuned the adaptive algorithm; see the appendix for a detailed explanation). 
\par
We note that the $X$ parameters are not arbitrary. Under the assumption that the errors have zero mean, $\hat{\mu}_X=130.757$ is just the observed sample mean, and $\hat{\sigma}_X=17.528$ is
\[\hat{\sigma}^2_X=\frac{1}{p}\Big\{
\frac{p\sum_{j=1}^n(\bar{Y}_{j.}-\bar{Y})^2}{n-1}-
\hat{\sigma}_\epsilon^2\Big\},\]
where $\bar{Y}=\frac{1}{n}\sum_{j=1}^n\bar{Y}_{j.}$. Table~\ref{tab:frag_sim} presents the results of $1000$ simulations which were carried out with a sample size of $n=500$ and contamination of about $50\%$ ($9.206/17.528$). These can be compared to the results for estimating $N(0,1)$ under $N(0,0.5^2)$ error in Table~\ref{tab:est_normal2}.
\begin{table}[h]		
\caption{Empirical RMSE and bias$\times$10 (in parentheses) for estimating $N(130.757,17.528^2)$ under $N(0,9.206^2)$ error.}																					
\label{tab:frag_sim}																					
\begin{center}																					
\begin{tabular}{lccccc}																					
\hline	
\hline																				
\multicolumn{1}{l}{}&																					
\multicolumn{5}{c}{$F_X(x_0)$}\label{fig:framingham}\\																					
\hline																					
Estimator&0.1&0.25&0.5&0.75&0.9\\																					
\hline			
\\																		
Adaptive	&	0.017	(0.088)&	0.022	(0.117)&	0.017	(0.007)&	0.022	(0.116)&	0.017	(0.080)	\\
SIMEX	&	0.019	(0.000)&	0.026	(0.005)&	0.029	(0.003)&	0.025	(0.003)&	0.019	(0.005)		\\
Naive	&	0.021	(0.148)&	0.024	(0.131)&	0.022	(0.002)&	0.024	(0.132)&	0.021	(0.153)	\\\\
\hline																					
\end{tabular}																					
\end{center}																					
\end{table}

\par
We see that for the specific parametric set up here, the adaptive estimator is uniformly better than the SIMEX and naive estimators in terms of root mean square error. The large $\sigma_X$ in this case indicates the smoothness of the $X$ distribution. If we consider theoretical aspects of these methods, then the good theoretical properties of the adaptive estimator described above, guarantee that in the minimax sense, no other estimator can do better over the class of finite smoothness distributions. 
\subsection{Statistical inference}
When estimating a disease prevalence, an applied statistician may not be satisfied with only pointwise properties of a new method, no matter how good they are. Thus, the next natural step would be to discuss the accuracy of the adaptive estimator and provide interval estimation. However, it is a known fact that confidence bands cannot adapt to the smoothness of the unknown function $F_X$ (see \citeasnoun{low}). One possibility would be to use bootstrap confidence intervals but in our case they require heavy computational efforts with no underlying theory to justify them. For practical implementation we suggest using the following approach. 
\par 
Let $\tau=\sqrt{\frac{F_Y(x_0)(1-F_Y(x_0))}{n}}$ and consider the following asymptotically based $1-\alpha$ confidence interval for $F_Y(x_0)$,
\begin{eqnarray}\label{eq:ci1}
1-\alpha=P\Big\{
\hat{F}_Y(x_0)-z_{1-\alpha/2}\tau\leq F_Y(x_0)
\leq\hat{F}_Y(x_0)+z_{1-\alpha/2}\tau\Big\},
\end{eqnarray}
where $z_{1-\alpha/2}$ is the $1-\alpha/2$ quantile of the normal distribution and $\hat{F}_Y(x_0)$ is the empirical distribution function.
Now let us look at the right hand side of the interval in (\ref{eq:ci1}) and note that 
\begin{eqnarray*}
&&P\Big\{F_Y(x_0)\leq\hat{F}_Y(x_0)+z_{1-\alpha/2}\tau)\Big\}
\\&&=P\Big\{F_X(x_0)\leq\hat{F}_Y(x_0)+F_X(x_0)-F_Y(x_0)+z_{1-\alpha/2}\tau)\Big\}
\\&&\leq P\Big\{F_X(x_0)\leq\hat{F}_Y(x_0)+|F_X(x_0)-F_Y(x_0)|+z_{1-\alpha/2}\tau)\Big\}.
\end{eqnarray*}
Applying the same argument to the left hand side of the interval in (\ref{eq:ci1}) we obtain 
\begin{eqnarray}\label{eq:ci2}
P\Big\{F_X(x_0)\in\big\{\hat{F}_Y(x_0)\pm\big[|F_X(x_0)-F_Y(x_0)|+z_{1-\alpha}\tau\big]\big\}\Big\}\geq 1-\alpha.
\end{eqnarray}
Note that when there is no measurement error $F_X(x_0)=F_Y(x_0)$ and the interval (\ref{eq:ci2}) reduces to that in (\ref{eq:ci1}). If the error is moderate, then we expect that the interval (\ref{eq:ci2}) would be somewhat conservative but still reasonable. However, this interval is based on unknown quantities and can not be practically applied. Therefore, we use its empirical counterpart by plugging in the estimators for $\tau$ and $F_X(x_0)$ as follows:
\begin{eqnarray}\label{eq:conint}
CI[F_X(x_0)]:=\big\{\hat{F}_Y(x_0)\pm\big[|\hat{F}_A(x_0)-\hat{F}_Y(x_0)|+z_{1-\alpha}\hat{\tau}\big]\big\},
\end{eqnarray}
where $\hat{F}_A(x_0)$ stands for the adaptive estimator, $\hat{F}_Y(x_0)$ for the empirical distribution function and 
\[
\hat{\tau}=\sqrt{\frac{\hat{F}_Y(x_0)(1-\hat{F}_Y(x_0))}{n}}.
\]
\par 
Simulation results presented in Table~\ref{tab:reallifecoverage} indicate that the observed coverage of this interval for $\alpha=0.05$ was close to the nominal $95\%$ level. 
\begin{table}[ht]
\caption{Empirical coverage intervals and probabilities for estimating $N(130.757,17.528^2)$ under $N(0,9.206^2)$ error based on $1000$ samples of size $n=500$. Here $\alpha=0.05$. The intervals and widths are averages over the $1000$ samples.}
\label{tab:reallifecoverage}
\begin{center}
\begin{tabular}{lccccc}
\hline	\hline																				
\multicolumn{1}{l}{}&																					
\multicolumn{5}{c}{$F_X(x_0)$}\\																					
\hline		
&0.1&0.25&0.5&0.75&0.9\\
\hline
\\
Interval&[0.08,0.15]&[0.22,0.30]&[0.45,0.55]&[0.70,0.78]&[0.85,0.92]\\
Width&0.07&	0.09&	0.1&	0.09&	0.07\\
Coverage&93.6\%&94.1\%&98.5\%&94.1\%&93.5\%\\\\
\hline
\end{tabular}
\end{center}
\end{table}
\subsection{Estimation in the data example}
We now turn to estimation of the hypertension prevalence.
Here we assume that the measurement error is normally distributed, but unlike the above simulation study, no distributional assumption is made about $X$.
\par 
The naive estimator in our case is $1-\hat{F}_Y(140)=0.225$ while the SIMEX estimator is $1-\hat{F}_S(140)=0.184$. The adaptive estimator is $1-\hat{F}_A(140)=0.21$ and the interval given by (\ref{eq:conint}) is $[0.19,0.26]$ (which does not include the SIMEX estimator). 
\par
\begin{figure}
\centerline{\includegraphics[width=5in]{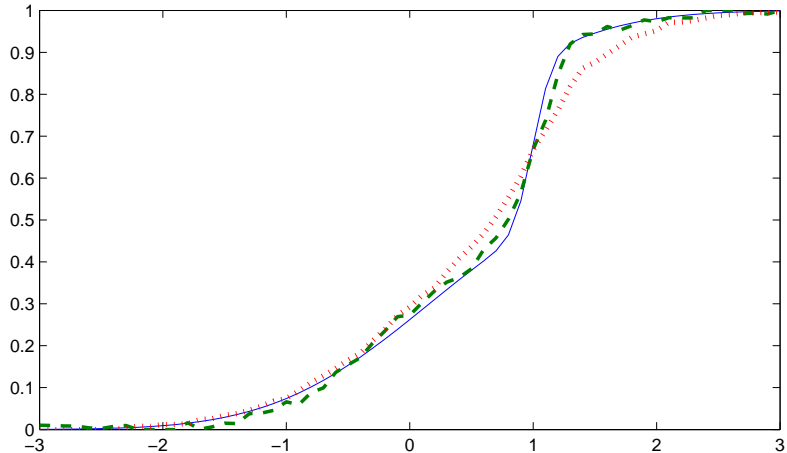}}
\caption{One realization of estimating normal mixture $N(0.15827,1)+N(1,0.1225^2)$ under Laplace error with scale $1/(2\sqrt{2})$. Sample size $n=500$. The solid line, dashed line, and dotted line correspond
to the true distribution, adaptive and naive estimators respectively.}
\label{fig:realiz500}\end{figure}

 
The fact that both the naive and the adaptive estimator yield similar estimation results may give the wrong impression that these methods behave the same. One then may prefer to use the naive estimator since it is more straightforward to implement. However, although in the example above the results are similar, in other examples they may differ substantially. This depends on the estimated distribution which of course is not known to us. This is well illustrated by Figure~\ref{fig:realiz500} where we see one realization of estimating the normal mixture $N(0.15827,1)+N(1,0.1225^2)$ under Laplace error (with scale $1/(2\sqrt{2})$) for $n=500$. The adaptive estimator adapts to the underlying smoothness of the unknown normal mixture all over its quantiles. However, the naive estimator behaves nicely in places where the underlying distribution is smooth but worse when it is not. Thus, the adaptive methods guarantee that in general we do better although in particular cases we may not. 

\subsection{Sensitivity Analysis.}
In our example we used an estimate for the measurement error variance and not the unknown true value. In this case a sensitivity analysis of our results to different values of the error variance would be informative. Under the assumption that both the estimated distribution and the error distribution are normally distributed, \citeasnoun{searle} provide an unbiased estimate for the variance of $\hat{\sigma}_\epsilon^2$ which is
\[\hat{
{\rm var}}(\hat{\sigma}_\epsilon^2)=\frac{2\hat{\sigma}_\epsilon^4}{n(p-1)+2}.\]
Under the assumption that the error is normally distributed, we calculated the adaptive estimator for a set of ten (equal spaced) values of $\sigma_\epsilon$ ranging from $\hat{\sigma}_\epsilon^2-2\sqrt{\hat{{\rm var}}(\hat{\sigma}_\epsilon^2)}$ to  $\hat{\sigma}_\epsilon^2+2\sqrt{\hat{{\rm var}}(\hat{\sigma}_\epsilon^2)}$. Specifically, in our case we have $\sqrt{\hat{{\rm var}}(\hat{\sigma}_\epsilon^2)}=2.981$ and the different estimates are given in Table~\ref{tab:sens}.
\begin{table}[ht]
\caption{Sensitivity analysis for the adaptive estimator.}
\label{tab:sens}
\begin{center}
\begin{tabular}{ccc}
\hline	\hline																				
\multicolumn{1}{c}{$\sigma_\epsilon^2$}&																					
\multicolumn{1}{c}{Estimator}&																					
\multicolumn{1}{c}{Interval}\\																					
	
\hline
\\
78.793&0.209&[0.19,0.26]\\
80.118&0.209&[0.19,0.26]\\
81.443&0.209&[0.19,0.26]\\
82.767&0.210&[0.19,0.26]\\
84.092&0.210&[0.19,0.26]\\
85.417&0.210&[0.19,0.26]\\
86.742&0.211&[0.19,0.26]\\
88.067&0.204&[0.18,0.27]\\
89.391&0.205&[0.18,0.27]\\
90.716&0.205&[0.18,0.27]\\
\\
\hline
\end{tabular}
\end{center}
\end{table}
We see that the adaptive estimator stays very close to its initial value of $0.21$ and is smaller than the naive estimate in all cases. The interval's upper and lower values (and width) show very little change. Thus, the adaptive estimator seems in our example to be robust to the fact that we estimate the measurement error variance.  
\section{Discussion}\label{sec:dis}
The problem of pointwise estimation of a distribution function in measurement error models was studied. Our estimation method was based on a direct inversion formula for the distribution function. This method was shown to be minimax optimal for ordinary smooth error distributions in \citeasnoun{DGJ}. We have shown here that it is also minimax optimal for supersmooth error distributions and provided an adaptive version for this case. In particular, we have shown that there is no payment in the rate of convergence when adapting under supersmooth error distribution.   
\par 
An extensive simulation study was carried out in order to study finite sample properties of the aformentioned method. The adaptive estimator performs well in different estimation setups and seems to be the only reasonable estimator when the error distribution is not symmetric with non-zero expectation. 
\par 
The application of our method to a real data example was examined and different practical aspects were explored. In particular, the data we considered are based on repeated measures and the estimation of the error variance was taken into account by modifying our estimation procedure to allow for the estimation of this parameter. The theoretical consequences of doing so are not yet known but simulation results are promising and in our particular example the adaptive estimator seems to be robust. The use of different assumptions for the error distribution can results in different estimates. In our data example we assumed that the measurement error is normally distributed. If the underlying error distribution is Laplace then the adaptive estimator is $1-\hat{F}_A(140)=0.189$ while if the error distribution is Gamma with shape parameter two and relocated to have zero expectation, then the adaptive estimator is $1-\hat{F}_A(140)=0.178$. 
\par
This emphasizes the importance of developing methods without assuming a distributional form for the error. This estimation problem has been thoroughly studied for density deconvolution (see \citeasnoun{johannes} and references therein) and similar paths may be taken for the distribution case. For instance, assuming that we have at hand an additional sample of directly observed measurement errors we can estimate the characteristic function $\phi_\epsilon$ by its empirical version. In general, this approach may lead to instable results and it is preferable to use a modified estimator in which only "good" estimates of $\phi_\epsilon$ are taken into account. This method was shown to be minimax optimal for density deconvolution in \citeasnoun{Neumann} and we are able to show similar theoretical results for distribution deconvolution. However, as already mentioned, this is not enough for practical considerations and an adaptive version of the estimator is required. The study of this problem is beyond the scope of this paper and will be considered elsewhere.
\section*{Acknowledgment}
The first author was supported by BSF grant 2006075. The authors thank Alexander Goldenshluger for helpful discussions.
\par 
The authors are grateful to the Associate Editor and one anonymous referee for
careful reading and useful remarks that led to substantial improvements in the presentation.
\section*{Appendix}
\subsection{Proof of Theorem 1}
The proof is based on the standard bias-variance decomposition
\begin{eqnarray*}\label{eq:b-v}
 E |\hat{F}_\lambda(x_0)-F_X(x_0)|^2 &=& | E \hat{F}_\lambda(x_0)
-F_X(x_0)|^2 + E |\hat{F}_\lambda(x_0)- E\hat{F}_\lambda(x_0)|^2 
\nonumber
\\
&=:&
B_\lambda^2(F_X;x_0) + {\rm var}\{\hat{F}_\lambda(x_0)\}.
\end{eqnarray*}
\subsubsection{Bounding the bias}
Note that
\begin{eqnarray*}
\rE[\hat{F}_{\lambda}]= \frac{1}{2}-
\frac{1}{\pi}\int_0^\lambda \omega^{-1} \Im\{e^{-i\omega
x_0}\phi_X(\omega)\}d\omega.
\end{eqnarray*}
Therefore it follows from (\ref{eq:inversion}) that
\begin{eqnarray*}
B_\lambda(F_X;x_0)&=&
\bigg|\frac{1}{\pi}\int_\lambda^\infty\omega^{-1} \Im(e^{-i\omega
x_0}\phi_X(\omega)) d\omega\bigg|
\;\leq\;\frac{1}{\pi}\int_\lambda^\infty
\omega^{-1}|\phi_X(\omega)| d\omega.
\end{eqnarray*}
For $\alpha\geq 0$  using  the Cauchy--Schwarz inequality we obtain
\begin{eqnarray*}
&& B_\lambda(F_X;x_0)
\;\leq\;
\frac{1}{\pi}\int_\lambda^\infty\frac{|\phi_X(\omega)|}
{\omega}d\omega
\\
&&
\;\leq\;
\frac{1}{\pi}
\bigg(\int_\lambda^\infty|\phi_X(\omega)|^2
(1+ \omega^{2})^\alpha d\omega\bigg)^{1/2}
\bigg(\int_\lambda^\infty \frac{1}{\omega^{2\alpha+2}}d\omega\bigg)^{1/2}
\; \leq\; 
\sqrt{\frac{2}{\pi}}\,L\,
\frac{\lambda^{-\alpha-1/2}}{\sqrt{2\alpha+1}}.
\end{eqnarray*}
If $\alpha\in (-1/2, 0)$ then for any $\lambda\geq 1$
\[
 B_\lambda(F_X;x_0) \;\leq\; \sqrt{\frac{2}{\pi}}L
\Big(\int_\lambda^\infty \frac{1+\omega^{-2\alpha}}{\omega^2}\,d \omega
\Big)^{1/2} \;\leq\;
\sqrt{\frac{2}{\pi}}L \big[1+(2\alpha+1)^{-1/2}\big] \lambda^{-\alpha-1/2}.
\]
Combining the two bounds we obtain the following bound for bias of the estimator,
\begin{equation}\label{eq:bias}
\sup_{F_X\in \cS_\alpha(L)}
B_\lambda(F_X; x_0) \;\leq\;
K_0
L \lambda^{-\alpha - 1/2},\;\;
K_0:=\sqrt{2/\pi}[1+ (2\alpha+1)^{-1/2}].
\end{equation}
\subsubsection{Bounding the variance}
The following lemma will be used in the sequel.
\begin{lemma}\label{lem:image_bound}
For any $\omega, \mu\in \rR$ and $x_0\in \rR$ one has
\begin{eqnarray*}
\bigg|\int_{-\infty}^\infty
\Im\bigg\{\frac{e^{i\omega(y-x_0)}}{\phi_\epsilon(\omega)}\bigg\}
\Im\bigg\{\frac{e^{i\mu(y-x_0)}}{\phi_\epsilon(\mu)}\bigg\}f_Y(y)d y\bigg|
\;\leq\;
\frac{|\phi_Y(\omega-\mu)| +|\phi_Y(\omega+\mu)|}
{2|\phi_\epsilon(\omega)|\,|\phi_\epsilon(\mu)|}~.
\end{eqnarray*}
\end{lemma}
\pr
Using (\ref{eq:Im}) we have
\begin{eqnarray*}
&& |\phi_\epsilon(\omega)|^2 |\phi_\epsilon(\mu)|^2
\Im\bigg\{\frac{e^{i\omega(y-x_0)}}{\phi_\epsilon(\omega)}\bigg\}
\Im\bigg\{\frac{e^{i\mu(y-x_0)}}{\phi_\epsilon(\mu)}\bigg\}
\\
&&\;=\;
\int_{-\infty}^\infty\int_{-\infty}^\infty
\sin\{\omega(y-x_0-u)\} \sin\{\mu(y-x_0-v)\} f_\epsilon(u) f_\epsilon(v)
\,d u\, d v
\\
&&\;=\;
\frac{1}{2}
\int_{-\infty}^\infty\int_{-\infty}^\infty
\cos\{(\omega-\mu)(y-x_0)-\omega u+\mu v\} f_\epsilon(u) f_\epsilon(v) \,d u\,d v
\\
&& \hspace{10mm}-
\frac{1}{2}
\int_{-\infty}^\infty\int_{-\infty}^\infty
\cos\{(\omega+\mu)(y-x_0)-\omega u -\mu v\} f_\epsilon(u) f_\epsilon(v)
\,d u\,d v
\\
&&\;=\;\frac{1}{2}
\Re \int_{-\infty}^\infty
\int_{-\infty}^\infty e^{-i\omega u}\Big[ e^{i(\omega-\mu)(y-x_0)}e^{i\mu v}
-  e^{i(\omega+\mu)(y-x_0)}e^{-i\mu v}\Big] f_\epsilon(u) f_\epsilon(v) \,d u
\,d v.
\end{eqnarray*}
Multiplying the last expression by $f_Y(y)$, integrating over $y$ and using
the Fubini theorem we obtain
\begin{eqnarray*}
&& |\phi_\epsilon(\omega)|^2|\phi_\epsilon(\mu)|^2 \int_{-\infty}^\infty
\Im\bigg\{\frac{e^{i\omega(y-x_0)}}{\phi_\epsilon(\omega)}\bigg\}
\Im\bigg\{\frac{e^{i\mu(y-x_0)}}{\phi_\epsilon(\mu)}\bigg\}f_Y(y)d y
\\
&&\;=\;\;\frac{1}{2}
\Re\int_{-\infty}^\infty
\int_{-\infty}^\infty e^{-i\omega u}\Big[
\phi_Y(\omega-\mu) e^{-i(\omega-\mu)x_0} e^{i\mu v} 
\\
&&\hspace{40mm}
-\;
\phi_Y(\omega+\mu) e^{-i(\omega+\mu)x_0}e^{-i\mu v}\Big]f_\epsilon(u)
f_\epsilon(v) d u\,d v
\\
&&\;=\;\;\frac{1}{2}\Re\,\Big\{
\overline{\phi_\epsilon(\omega)}\;\Big[\phi_\epsilon(\mu)
\phi_Y(\omega-\mu) e^{-i(\omega-\mu)x_0}
- \overline{\phi_\epsilon(\mu)}\;\phi_Y(\omega+\mu)
e^{-i(\omega+\mu)x_0}\Big]
\Big\}.
\end{eqnarray*}
The result of the lemma immediately follows from the last relation.
\epr
By definition of $\hat{F}_\lambda$ we can bound the variance of the estimator by the second moment as follows:
\begin{eqnarray*}
 {\rm var}\{\hat{F}_\lambda\} \;\leq\; \frac{1}{n}
\rE \bigg(\frac{1}{\pi}\int_0^\lambda \frac{1}{\omega} \Im
\bigg\{\frac{e^{i\omega(Y_j-x_0)}}
{\phi_\epsilon(\omega)}\bigg\}d \omega \bigg)^2.
\end{eqnarray*}
Let 
\begin{eqnarray}\label{eq:omega1}
\omega_1:=\min\{\omega_0,(2b_\epsilon)^{-1/\tau}\},
\end{eqnarray}
where $\omega_0$ and $b_\epsilon$ are defined in Assumption~\ref{as:local}.
Then we can write
\begin{gather}
{\rm var}\{\hat{F}_\lambda\}
 \leq \frac{2}{\pi^2n} \bigg\{\rE \Big(\int_0^{\omega_1} \frac{1}{\omega} \Im
\Big\{\frac{e^{i\omega(Y_j-x_0)}}
{\phi_\epsilon(\omega)}\Big\}d \omega \Big)^2 +
\rE \Big(\int_{\omega_1}^\lambda \frac{1}{\omega} \Im
\Big\{\frac{e^{i\omega(Y_j-x_0)}}
{\phi_\epsilon(\omega)}\Big\}d \omega \Big)^2
\bigg\}
\nonumber
\\
=: \frac{2}{\pi^2 n} (I_1 +I_2),
\label{eq:I1+I2}
\end{gather}
and we bound $I_1$ and $I_2$ separately.
\par
1$^0$. We begin with bounding $I_1$. First note that
\begin{eqnarray}
 \Im\big\{\phi_\epsilon^{-1}(\omega)e^{i\omega(y-x_0)}\big\}
&=& |\phi_\epsilon(\omega)|^{-2} \Im\Big\{e^{i\omega(y-x_0)}
\overline{\phi_\epsilon(\omega)}\Big\}
\nonumber
\\
&=&
|\phi_\epsilon(\omega)|^{-2}
\int_{-\infty}^\infty \sin\{\omega(y-x_0-u)\} f_\epsilon(u) d u.
\label{eq:Im}
\end{eqnarray}
Therefore
\[
 I_1 =\rE\bigg(\int_0^{\omega_1}
\int_{-\infty}^\infty\frac{
\sin\{\omega(y-x_0-u)\}}
{|\phi_\epsilon(\omega)|^2\omega}f_\epsilon(u)\,d u\,d\omega\bigg)^2
\;=:\;\int_{-\infty}^\infty [\bar{I}_1(y)]^2 f_Y(y) d y.
\]
First, observe that $|\phi_\epsilon(\omega)|^{-2}=1+r(\omega)$, where
$r(\omega):=\sum_{k=1}^\infty (|\phi_\epsilon(\omega)|^2-1)^k$.
In addition, by Assumption~\ref{as:local}, $|r(\omega)|\leq \sum_{k=1}^\infty (2b_\epsilon |\omega|^\tau)^k$ for all
$|\omega|\leq \omega_0$.
Hence
\begin{eqnarray*}
\Big|\int_0^{\omega_1}
\frac{\sin\{\omega(y-x_0-u)\}}{|\phi_\epsilon(\omega)|^2\omega} d\omega\Big|
&\leq&  \Big|\int_0^{\omega_1}
\frac{\sin\{\omega(y-x_0-u)\}}{\omega} d\omega\Big|
\;
+ \;\int_0^{\omega_1}
\frac{|r(\omega)|}{\omega} d\omega
\\
&\leq& 2 + \sum_{k=1}^\infty \frac{(2b_\epsilon \omega_1^{\tau})^k}{\tau k}\;\leq\;
2+\tau^{-1},
\end{eqnarray*}
where we have used the fact that $\sup_{x>0}\int_0^x t^{-1}\,\sin t\,d t <1.85195$ (see \citeasnoun{kawata}), the above upper bound
on $|r(\omega)|$ and the definition of $\omega_1$ in (\ref{eq:omega1}).
Therefore, by Fubini's  and dominated convergence theorems , we get
$|\bar{I}_1(y)|\leq (2+\tau^{-1})$  for all $y$,
which, in turn, implies that
\begin{equation}\label{eq:I-11}
 I_1 \;\leq \; [2+(1/\tau)]^2.
\end{equation}
\par
2$^0$. Now we bound $I_2$.
We have
\begin{eqnarray*}
I_2&=&\int_{\omega_1}^\lambda\int_{\omega_1}^\lambda
\frac{1}{\omega\mu}\bigg[
\int_{-\infty}^\infty
\Im\Big\{\frac{e^{i\omega(y-x_0)}}{\phi_\epsilon(\omega)}\Big\}
\Im\Big\{\frac{e^{i\mu(y-x_0)}}{\phi_\epsilon(\mu)}\Big\}
f_Y(y)d y\bigg]d\omega d\mu.
\end{eqnarray*}
Lemma \ref{lem:image_bound} implies that
\begin{eqnarray}
I_2&\leq&\int_{\omega_1}^\lambda\int_{\omega_1}^\lambda
\frac{|\phi_Y(\omega-\mu)|}
{2\omega\mu|\phi_\epsilon(\omega)|\,|\phi_\epsilon(\mu)|}
d\omega d\mu
+\int_{\omega_1}^\lambda\int_{\omega_1}^\lambda
\frac{|\phi_Y(\omega+\mu)|}
{2\omega\mu |\phi_\epsilon(\omega)|\,|\phi_\epsilon(\mu)|}
d\omega d\mu
\nonumber
\\
&=:& I_{2}^{(1)} + I_{2}^{(2)}.
\nonumber
\end{eqnarray}
Using the Cauchy-Schwarz inequality we have
\begin{equation}\label{eq:I21}
I_2^{(1)}\leq \frac{1}{2}
\bigg(\int_{\omega_1}^\lambda\int_{\omega_1}^\lambda
\frac{|\phi_Y(\omega-\mu)|}
{|\phi_\epsilon(\omega)|^2\omega^2}
d\omega d\mu\bigg)^{1/2}
\bigg(\int_{\omega_1}^\lambda\int_{\omega_1}^\lambda
\frac{|\phi_Y(\omega-\mu)|}
{|\phi_\epsilon(\mu)|^2\mu^2}
d\omega d\mu\bigg)^{1/2}.
\end{equation}
Because
$\phi_Y(\omega)=\phi_X(\omega) \phi_\epsilon(\omega)$ and $|\phi_X(\omega)|\leq 1$
we have for  any $\omega\in [\omega_1, \lambda]$
\begin{eqnarray*}
 \int_{\omega_1}^\lambda |\phi_Y(\omega-\mu)| d \mu &\leq&
\int_{-\lambda}^\lambda |\phi_\epsilon (\omega)|\,|\phi_X(\omega)|\,d \omega
\leq
c_1\int_{-\lambda}^\lambda e^{-\gamma|\omega|^\beta}\;d \omega,
\end{eqnarray*}
where we have used the upper bound
in  Assumption~\ref{as:global}. Substituting $t=\gamma\omega^\beta$ we see that
\begin{eqnarray}\label{eq:boundnominator}
\int_{\omega_1}^\lambda |\phi_Y(\omega-\mu)| d \mu\leq
\frac{2c_1}{\gamma^{1/\beta}\beta}\int_0^{\gamma\lambda^\beta}
e^{-t}t^{1/\beta-1}d t
\leq\frac{2c_1\Gamma(1/\beta)}{\gamma^{1/\beta}\beta},
\end{eqnarray}
where $\Gamma(z)$ is the gamma function $\Gamma(z)=\int_0^\infty e^{-t}t^{z-1}dt$. 
Now, using the lower bound in Assumption~\ref{as:global} we obtain
\begin{eqnarray*}
\int_{\omega_1}^\lambda
\frac{1}{\omega^2|\phi_\epsilon(\omega)|^2}d\omega&\leq&
\frac{1}{\omega_1^2c_0^2}
\int_{\omega_1}^\lambda e^{2\gamma|\omega|^\beta}d\omega
\leq\frac{\lambda e^{2\gamma\lambda^\beta}}{\omega_1^2c_0^2}.
\end{eqnarray*}
The last bound together with (\ref{eq:boundnominator}) and (\ref{eq:I21}) leads to 
\begin{eqnarray*}
I_2^{(1)}\leq\frac{2c_1\Gamma(1/\beta)}{\gamma^{1/\beta}\beta\omega_1^2c_0^2}\lambda e^{2\gamma\lambda^\beta},
\end{eqnarray*}
which holds also for $I_2^{(2)}$. Therefore we conclude that 
\[I_2\leq\frac{4c_1\Gamma(1/\beta)}{\gamma^{1/\beta}\beta\omega_1^2c_0^2}\lambda e^{2\gamma\lambda^\beta}.\]
\par
3$^0$. We now combine the bounds for $I_1$ given in (\ref{eq:I-11}) and the bound for $I_2$ above together with (\ref{eq:I1+I2}) to get
\begin{eqnarray}\label{eq:varbound}
{\rm var}\{\hat{F}_\lambda(x_0)\}\leq 
\frac{2}{\pi^2n}\Big\{
[2+(1/\tau)]^2+
\frac{4c_1\Gamma(1/\beta)}{\gamma^{1/\beta}\beta\omega_1^2c_0^2}\lambda e^{2\gamma\lambda^\beta}
\Big\}.
\end{eqnarray}
\subsubsection{Finding the optimal bandwidth}
Recall the definition for $\omega_1$ given in (\ref{eq:omega1}), let $\Gamma(z)$ be the gamma function $\Gamma(z)=\int_0^\infty e^{-t}t^{z-1}dt$ and define
\begin{eqnarray}\label{eq:ceps}
c_\epsilon:=\frac{2}{\pi^2}\Big\{
[2+(1/\tau)]^2+
\frac{4c_1\Gamma(1/\beta)}{\gamma^{1/\beta}\beta\omega_1^2c_0^2}\Big\}.
\end{eqnarray}
The bound in (\ref{eq:varbound}) implies that for $\lambda\geq 1$ we have ${\rm var}\{\hat{F}_\lambda(x_0)\}\leq c_\epsilon\lambda e^{2\gamma\lambda^\beta}n^{-1}$. 
\par 
We now wish to balance the squared bias with the variance by solving for $\lambda$ the equation
\begin{eqnarray}\label{eq:balance}
c_\epsilon\lambda e^{2\gamma\lambda^\beta}n^{-1}=K_0^2L^2 \lambda^{-2\alpha - 1},
\end{eqnarray}
where the constant $K_0$ is given in (\ref{eq:bias}). That yields
\[
\lambda^*=\Big\{
\frac{\ln n}{2\gamma}-\frac{\ln c_\epsilon+(2\alpha+2)\ln\lambda^*-2\ln(K_0L)}
{2\gamma}
\Big\}^{1/\beta}.
\]
Now note that for large enough $n$ we have $\lambda^*\leq\big[\ln n/(2\gamma)\big]^{1/\beta}$, thus
\[
\lambda^*\geq\Big\{\frac{\ln n}{2\gamma}-\frac{\ln c_\epsilon+\frac{(2\alpha+2)}{\beta}\ln\Big(\frac{\ln n}{2\gamma}\Big)-2\ln(K_0L)}
{2\gamma}
\Big\}^{1/\beta}=\lambda_\star
\]
as given in the theorem. Indeed, plugging $\lambda_\star$ in (\ref{eq:balance}) and noting that for large enough $n$
\[\Big(\frac{\ln n}{4\gamma}\Big)^{1/\beta}\leq\lambda_\star\leq\Big(\frac{\ln n}{2\gamma}\Big)^{1/\beta},\]
the theorem follows. 
\epr
\subsection{Proof of Theorem 2}
The idea is to choose $\lambda$ smaller then the optimal $\lambda_\star$ so that it will make the bias dominant. To this end, note that for large enough $n$ 
\[
\frac{2\alpha+2}{\beta}\leq\ln\Big(\frac{\ln n}{2\gamma}\Big),
\]
which implies that
\[
\hat{\lambda}=\Big\{\frac{\ln n}{2\gamma}-\frac{\ln c_\epsilon+\Big[\ln\Big(\frac{\ln n}{2\gamma}\Big)\Big]^2}
{2\gamma}
\Big\}^{1/\beta}
\leq\Big\{\frac{\ln n}{2\gamma}-\frac{\ln c_\epsilon+\frac{(2\alpha+2)}{\beta}\ln\Big(\frac{\ln n}{2\gamma}\Big)-2\ln(K_0L)}
{2\gamma}
\Big\}^{1/\beta}=\lambda_\star.
\]
Therefore $e^{2\gamma\hat{\lambda}^\beta}\leq e^{2\gamma\lambda_\star^\beta}$. Finally, here also for large enough $n$ we have $\hat{\lambda}\geq\big[(\ln n)/(4\gamma)\big]^{1/\beta}$, thus, plugging back in (\ref{eq:balance}) these bounds for $\hat{\lambda}$ the theorem follows. 
\epr

\subsection{Tuning of the adaptive algorithm}
Here we describe in detail the tuning of the adaptive algorithm. As already mentioned above, theoretically, $K_\epsilon$ depends only on the error distribution which is assumed to be completely known, and its exact value can be computed for any error distribution explicitly (see \citeasnoun{DGJ}). However, numerical experience suggests that the theoretical value of $K_\epsilon$ is too conservative. Thus, in practice we calibrated the adaptive algorithm as follows.
\par
We set $X$ to be standard normal, $\epsilon$ to be Laplace with standard deviation $\sigma_\epsilon$, $x_0$ is the value for which $F_X(x_0)=0.25$, and the sample size $n=2000$. The standard deviation of the measurement error takes the values $\sigma_\epsilon=0.05(0.1)0.95$. 
Let $\hat{\sigma}_\lambda$ be defined as in (\ref{eq:sigma_lambda}). For each $\sigma_\epsilon$ we estimated $F_X(x_0)$ using the interval 
\[\Big[\hat{F}_\lambda(x_0) - 
c_\epsilon\Big\{\frac{\ln(n)}{n}\Big\}^{1/2}\hat{\sigma}_\lambda,\;
\hat{F}_\lambda(x_0) + c_\epsilon\Big\{\frac{\ln(n)}{n}\Big\}^{1/2}\hat{\sigma}_\lambda\Big]\]
 for a set of different values of $c_\epsilon=0.01(0.02)10$. This procedure is repeated a hundred times and the value $c_\epsilon$ which minimized the empirical root mean square error of the adaptive estimator is chosen, and denoted by $c_{\sigma_\epsilon}$. This calculation was repeated fifty times which resulted in the fifty values $c_{\sigma_\epsilon,1},...,c_{\sigma_\epsilon,50}$. 
The mean of these values was taken and is denoted by $\bar{c}_{\sigma_\epsilon}$. This results in ten values of $\bar{c}_{\sigma_\epsilon}$ corresponding to the ten values of $\sigma_\epsilon$. Then a simple regression with the values of $\sigma_\epsilon$ as the independent variables, and those of $\bar{c}_{\sigma_\epsilon}$ as the dependent variable results in the rule $K_\epsilon:=\hat{\bar{c}}_{\sigma_\epsilon}=0.0275+0.3074\sigma_\epsilon$.
\par 
We note that the choice of $X$ to be standard normal and $F_X(x_0)=0.25$ in our calibration is arbitrary, at least theoretically. As mentioned above, the theoretical value of $K_\epsilon$ depends only on the error distribution. Indeed, calibration with different choices for the distribution of $X$ and the value of $x_0$ yielded similar results for a given error distribution.   
\par 
We further note that our study of the practical choice of $K_\epsilon$ is based on values of $\sigma_\epsilon$ smaller than one. If $\sigma_\epsilon$ is larger than one, we standardize the observed sample so that it will have zero mean and standard error of one. Then we use a standardized form of $\sigma_\epsilon$ in our procedure, i.e., the estimate $\sigma_\epsilon/\hat{\sigma}_Y$, where $\hat{\sigma}_Y$ is estimated from the observations.

\bibliographystyle{agsm}

\end{document}